**Spin and momentum of the light fields in inhomogeneous dispersive media with application to surface plasmon-polariton waves**

[1,2] Bekshaev A. Y. and [2,3] Bliokh K. Y.

[1] I.I. Mechnikov National University, Research Institute of Physics, Dvorianska 2, 65082, Odessa, Ukraine, e-mail: bekshaev@onu.edu.ua
[2] Center for Emergent Matter Science, RIKEN, Wako-shi, Saitama 351-0198, Japan
[3] Nonlinear Physics Centre, RSPE, The Australian National University, Canberra, Australia, e-mail: k.bliokh@gmail.com

**Abstract**

Following to the recently published approach [*Phys. Rev. Lett.* **119**, 073901 (2017); *New J. Phys*., **19**, 123014 (2017)], we refine and accomplish the general scheme for the unified description of the momentum and angular momentum in complex media. The equations for the canonical (orbital) and spin linear momenta, orbital and spin angular momenta in a lossless inhomogeneous dispersive medium are presented in the compact form analogous to the Brillouin's relation for the energy. The results are applied to the surface plasmon-polariton (SPP) field, and the microscopic calculations support the phenomenological expectations. The refined general scheme correctly describes the unusual SPP properties (transverse spin, magnetization momentum) and additionally predicts the singular momentum contribution sharply localized at the metal-dielectric interface, which is confirmed by the microscopic analysis. The results can be useful in optical systems employing the structured light, especially for microoptics, plasmophotonics, optical sorting and micromanipulation.

**Keywords**: electromagnetic momentum, angular momentum, dispersive media, spin-orbital decomposition, Abraham–Minkowski dilemma, surface plasmon-polariton

**PACS**: 42.25.Bs, 42.50.Tx, 42.50.Nn, 42.50.Wk, 03.50.De

## 1. Introduction

Properties of structured light fields attract growing attention during the past decade [1–3]. Such fields are necessary elements of multiple modern applications aimed at the optical trapping, sorting, delivering, selective treatment, positioning and other precise manipulations with extremely small quantities of matter [4–7]. In these areas, the optical field dynamical characteristics, i.e., first of all, spatial distribution of its energy, momentum and angular momentum (AM), play a crucial role, and their investigation is highly relevant. It is very important that the electromagnetic interactions in such systems normally develop on the highly inhomogeneous material background, which invokes the problem of "structured light in structured media".

Unfortunately, up to the recent time, the very instruments of the dynamical characteristics' description were well established only for the case when the field evolves in free space; in presence of continuous material media, even the introductory definitions of the field momentum become controversial and ambiguous. Over 100 years the debate continues between the Abraham and Minkowski momentum paradigms [8–11]; most reasonable "resolutions" of the dilemma find arguments supporting each side and treat both momenta as different physical quantities with their own scopes and abilities. However, the known analyses are mostly limited to the homogeneous media and plane-wave-like fields. Additional difficulties appear if the medium shows dispersion, i.e. the material parameters (permittivity and/or permeability) depend on frequency; the only field characteristic for which the dispersion can be taken into account in a regular and consistent way, is the energy whose density can be described by the famous Brillouin's formula [12,13].

Recently, Philbin and Allanson [14,15] have made a important advance and proposed a regular consistent way for description of the momentum and AM in dispersive media. But the genuine value of their approach becomes clear only in conjunction with the so called canonical decomposition of the field momentum when it is subdivided into the spin and canonical (orbital) components [16,17]. With further elaboration and microscopic substantiation, this approach resulted in the unified, compact and physically transparent expressions for the canonical linear momentum as well as orbital and spin AM of an optical field in dispersive inhomogeneous lossless media. The methods and results of [16,17] enabled to build the rigorous consistent theory of the surface plasmon-polariton (SPP), thoroughly analyze its non-trivial properties (e.g., the transverse spin), and predict novel phenomena (e.g., the SPP-induced magnetization of the media).

This work is aimed to further refinement of the recently developed methodology [16,17]. In particular, we enhance their general scheme to include the spin linear momentum whose description in dispersive media was previously omitted. As we will see, this enables to obtain the full set of instruments for the description and analysis of the electromagnetic momentum and AM in dispersive media, and to shed new light on some important results of the earlier works relating the SPP properties. In particular, we reveal some peculiar features of the momentum and spin distribution associated with near-surface contributions and the physically essential singular components of momentum sharply localized at the metal-dielectric interface.

The present consideration is essentially based on the materials of Refs. [16,17]; we not only employ their main ideas but, where possible, preserve their notations and terminology.

## 2. General overview of the dispersion-modified optical momentum description in media

In this paper, we deal with monochromatic fields in lossless dielectric media where the electric and magnetic vectors $\boldsymbol{\mathcal{E}}(\mathbf{r},t) = \mathrm{Re}\left[\mathbf{E}(\mathbf{r})e^{-i\omega t}\right]$ and $\boldsymbol{\mathcal{H}}(\mathbf{r},t) = \mathrm{Re}\left[\mathbf{H}(\mathbf{r})e^{-i\omega t}\right]$ obey the Maxwell equations

$$\nabla \cdot (\mu \mathbf{H}) = 0 , \quad \mu \mathbf{H} = -\frac{i}{k_0} \nabla \times \mathbf{E} ,$$

$$\nabla \cdot (\varepsilon \mathbf{E}) = 0 , \quad \varepsilon \mathbf{E} = \frac{i}{k_0} \nabla \times \mathbf{H} . \tag{1}$$

The medium is characterized by the real permittivity $\varepsilon = \varepsilon(\omega,\mathbf{r})$ and permeability $\mu = \mu(\omega,\mathbf{r})$ that may depend on coordinates (inhomogeneity) and on frequency (dispersion), $k_0 = \omega / c$ is the free-space wavenumber and $c$ is the vacuum light velocity. The only dynamical property of the field whose definition in such conditions is well established and free from controversies is the electromagnetic energy with the density described by the well-known Brillouin expression [12,13]:

$$\tilde{W} = \frac{g\omega}{2}\left( \tilde{\varepsilon}|\mathbf{E}|^2 + \tilde{\mu}|\mathbf{H}|^2 \right) , \tag{2}$$

$$\tilde{\varepsilon} = \varepsilon + \omega\frac{d\varepsilon}{d\omega} , \quad \tilde{\mu} = \mu + \omega\frac{d\mu}{d\omega} \tag{3}$$

where $g = (8\pi\omega)^{-1}$ (from now on, all dispersion-modified electromagnetic quantities are marked by tildes "~"). Note the neat and unified form of this expression which is valid for inhomogeneous media and differs from the dispersion-free formula [12,13] just by replacement $(\varepsilon,\mu) \to (\tilde{\varepsilon},\tilde{\mu})$ according to (3). Regrettably, there is no such a straight way for generalization of the field momentum and AM [16,17].

In the case of negligible dispersion, the Abraham and Minkowski momentum densities are given by [8–12]

$$\boldsymbol{\mathcal{P}}_A = gk_0 \,\mathrm{Re}\left(\mathbf{E}^* \times \mathbf{H}\right) , \tag{4}$$

$$\boldsymbol{\mathcal{P}}_M = gk_0\varepsilon\mu \,\mathrm{Re}\left(\mathbf{E}^* \times \mathbf{H}\right) . \tag{5}$$

These momenta are sometimes referred to as "kinetic" because they appear in the kinetic (symmetrical) energy-momentum tensor [18] of the electromagnetic field. Abstracting from the Abraham – Minkowski dilemma [8–11], both kinetic momenta meet difficulties in application to structured light fields [16,17]. Besides, corresponding AM densities

$$\boldsymbol{\mathcal{J}}_{A,M} = \mathbf{r} \times \boldsymbol{\mathcal{P}}_{A,M} \tag{6}$$

are "extrinsic" (depend on the choice of the coordinate origin), and the kinetic formalism based on expressions (4) or (5) cannot describe separate contributions of the spatial ("orbital") and polarization ("spin") degrees of freedom of light, which are intensively studied subjects of modern optics [3,19]. These drawbacks are partly eliminated in the "canonical" approach associated with the spin-orbital decomposition of the field AM [16–18,20–22]. This procedure manifests the especially favorable properties of the Minkowski momentum (5) which can be represented as

$$\mathcal{P}_M = \mathbf{P}_M + \mathbf{P}_M^S, \quad \mathbf{P}_M^S = \frac{1}{2}\nabla \times \mathbf{S}_M \tag{7}$$

where

$$\mathbf{P}_M = \frac{g}{2}\operatorname{Im}\left[\varepsilon\mathbf{E}^*\cdot(\nabla)\mathbf{E} + \mu\mathbf{H}^*\cdot(\nabla)\mathbf{H}\right] \tag{8}$$

is the "canonical" momentum, and

$$\mathbf{S}_M = \frac{g}{2}\operatorname{Im}\left(\varepsilon\mathbf{E}^* \times \mathbf{E} + \mu\mathbf{H}^* \times \mathbf{H}\right) \tag{9}$$

is the Minkowski spin density. The representation (7) – (9) is grounded on the Maxwell equations (1) and, remarkably, holds for arbitrary spatial-dependent $\varepsilon$ and $\mu$ (the similar operation with the Abraham momentum (4) is impossible because of emergence of additional terms owing to the medium inhomogeneity [16,17]). Accordingly, the Minkowski AM (6) can be reduced to

$$\mathcal{J}_M = \mathbf{r} \times \mathcal{P}_M = \mathbf{L}_M + \mathbf{S}_M \tag{10}$$

where

$$\mathbf{L}_M = \mathbf{r} \times \mathbf{P}_M \tag{11}$$

is the orbital AM which represents the extrinsic part of the total field AM, for which $\mathbf{S}_M$ (9) is the intrinsic part. Equation (10) is based on the non-local integral equality

$$\frac{1}{2}\int \mathbf{r} \times (\nabla \times \mathbf{S}_M)\, dV = \int \mathbf{S}_M\, dV = \int \mathbf{r} \times \mathbf{P}_M^S\, dV \tag{12}$$

valid for any fields properly vanishing at infinity. Eq. (12) expresses the general rule that in any electromagnetic field with inhomogeneous spin density $\mathbf{S}$, the corresponding *linear* spin momentum $\mathbf{P}^S$ exists with the density

$$\mathbf{P}^S = \frac{1}{2}\nabla \times \mathbf{S}\,, \tag{13}$$

and the second Eq. (7) is a special case of this rule.

Note that the canonical momentum (8) directly follows from the field Lagrangian as a conserved quantity via the Noether theorem [18,23,24]; this derivation leads to the non-symmetric (canonical) energy-momentum tensor. Then the linear spin momentum $\mathbf{P}^S$ appears as an auxiliary means for the tensor symmetrization by addition of the solenoidal momentum component [25]. However, recent studies (e.g., [16,17,20–22,26,27]), as well as Eqs. (7) – (5) and (13), disclose its deep physical meaning.

So far, the dispersion has been neglected in our reasoning. The important step to include the medium dispersion into the field momentum theory was made in Refs. [14,15] where the consideration is based on the field Lagrangian in a dispersive medium, and the momentum and AM expressions are derived via the Noether theorem. This mode of operation naturally has led to the Minkowski-based momentum representation and resulted in the following expressions for the field momentum and AM in a dispersive medium

$$\tilde{\mathcal{P}}_M^P = \mathcal{P}_M + \frac{g\omega}{2}\operatorname{Im}\left[\frac{d\varepsilon}{d\omega}\mathbf{E}^*\cdot(\nabla)\mathbf{E} + \frac{d\mu}{d\omega}\mathbf{H}^*\cdot(\nabla)\mathbf{H}\right], \tag{14}$$

$$\tilde{\mathcal{J}}_M = \mathbf{r}\times\tilde{\mathcal{P}}_M + \frac{g\omega}{2}\operatorname{Im}\left[\frac{d\varepsilon}{d\omega}\mathbf{E}^*\times\mathbf{E} + \frac{d\mu}{d\omega}\mathbf{H}^*\times\mathbf{H}\right] \tag{15}$$

(the superscript in Eq. (14) denotes that the dispersion correction is performed by means of the Philbin's procedure). However, the authors of [14,15] did not employ the spin-orbital decomposition (Eqs. (7) – (10)) without which the real meaning of their approach is underestimated. Indeed, with allowance for Eqs. (8), (9) and (11), the results (14) and (15) can be presented in the form

$$\tilde{\mathcal{P}}_M^P = \tilde{\mathbf{P}}_M + \mathbf{P}_M^S \tag{16}$$

where

$$\tilde{\mathbf{P}}_M = \frac{g}{2}\operatorname{Im}\left[\tilde{\varepsilon}\,\mathbf{E}^*\cdot(\nabla)\mathbf{E} + \tilde{\mu}\,\mathbf{H}^*\cdot(\nabla)\mathbf{H}\right] \tag{17}$$

and

$$\tilde{\mathcal{J}}_M = \tilde{\mathbf{L}}_M + \tilde{\mathbf{S}}_M \tag{18}$$

where

$$\tilde{\mathbf{S}}_M = \frac{g}{2}\operatorname{Im}\left(\tilde{\varepsilon}\,\mathbf{E}^*\times\mathbf{E} + \tilde{\mu}\,\mathbf{H}^*\times\mathbf{H}\right), \quad \tilde{\mathbf{L}}_M = \mathbf{r}\times\tilde{\mathbf{P}}_M\,. \tag{19}$$

Thus, the Philbin's transformations (14) and (15) provide explicit expressions for the canonical momentum $\tilde{\mathbf{P}}_M$ and spin AM $\tilde{\mathbf{S}}_M$ densities of the optical field in an inhomogeneous dispersive medium. Note the remarkably compact and unified character of the expressions (17) and (19): they merely reproduce the scheme in which the dispersion is taken into account in the Brillouin's formula (2) for the energy. In this form, equations for the field momentum and AM have been derived and used for the analysis of the SPP [16,17]. However, there are imperfections in the pattern described by Eqs. (16) – (19):

(i) in Eq. (16), the spin momentum $\mathbf{P}_M^S$ “does not feel” the dispersion and preserves the dispersion-free form (7), (9);

(ii) according to (13), the first Eq. (19) should entail the spin momentum expression that differs from that accepted in (16), namely

$$\tilde{\mathbf{P}}_M^S = \frac{1}{2}\nabla\times\tilde{\mathbf{S}}_M = \frac{g}{4}\nabla\times\mathrm{Im}\left(\tilde{\varepsilon}\,\mathbf{E}^*\times\mathbf{E}+\tilde{\mu}\,\mathbf{H}^*\times\mathbf{H}\right). \tag{20}$$

Obviously, the discrepancy appears because the procedure of [14] based on the Noether theorem gives the dispersion corrections for the conserved canonical momentum while the divergence-free spin momentum should be considered separately. Corresponding independent result for the spin momentum is just provided by (20), i.e. we can correct Eq. (16) into

$$\tilde{\mathcal{P}}_M = \tilde{\mathbf{P}}_M + \tilde{\mathbf{P}}_M^S \tag{16a}$$

which is equivalent to the following modification of the Philbin’s relation (14):

$$\tilde{\mathcal{P}}_M = \underbrace{\mathcal{P}_M + \frac{g\omega}{2}\mathrm{Im}\left[\frac{d\varepsilon}{d\omega}\mathbf{E}^*\cdot\left(\nabla\right)\mathbf{E}+\frac{d\mu}{d\omega}\mathbf{H}^*\cdot\left(\nabla\right)\mathbf{H}\right]}_{\tilde{\mathcal{P}}_M^P}$$

$$+\frac{g\omega}{4}\mathrm{Im}\left[\nabla\times\left(\frac{d\varepsilon}{d\omega}\mathbf{E}^*\times\mathbf{E}+\frac{d\mu}{d\omega}\mathbf{H}^*\times\mathbf{H}\right)\right]. \tag{14a}$$

Finally, the system of equations (16a), (20) and (17) – (19) completes the description of the field momentum and AM in inhomogeneous dispersive media. It is the main general statement of this work; some its consequences will be considered below.

## 3. Applications for a surface plasmon-polariton

Following to [16,17] we apply the derived equations to a very representative and non-trivial example of structured optical field in dispersive structured matter, which is supplied by the SPP wave at the metal-vacuum interface [4] (see Fig. 1a). The interface ($x = 0$ plane) separates the vacuum ($x > 0$, medium 1) and metal ($x < 0$, medium 2) half-spaces, making the system inhomogeneous; the SPP wave is the highly structured double-evanescent wave that exponentially decays on both sides from the interface and propagates along the $z$-axis with the well-defined wavevector $\mathbf{k}_p = k_p\overline{\mathbf{z}}$ (hereafter, $\overline{\mathbf{x}}$, $\overline{\mathbf{y}}$, and $\overline{\mathbf{z}}$ denote the unit vectors of the corresponding axes). The permittivity and permeability of the metal are described by the standard plasma model [4],

$$\mu = 1, \qquad \varepsilon(\omega) = 1 - \frac{\omega_p^2}{\omega^2}. \tag{21}$$

where

$$\omega_p^2 = \frac{4\pi n_0 e^2}{m} \tag{22}$$

is the plasma frequency, $n_0$ is the volume density of free electrons in the metal, $e < 0$ is the electron charge, and $m$ is the electron mass. Thus, the metal is a dispersive medium with $\tilde{\varepsilon} = 1 + \omega_p^2 / \omega^2 = 2 - \varepsilon \neq \varepsilon$, and the dispersion is crucial for the SPP properties. Even the existence of the SPP is conditioned by the frequency limit $\omega < \omega_p / \sqrt{2}$, that is, $\varepsilon < -1$ [4].

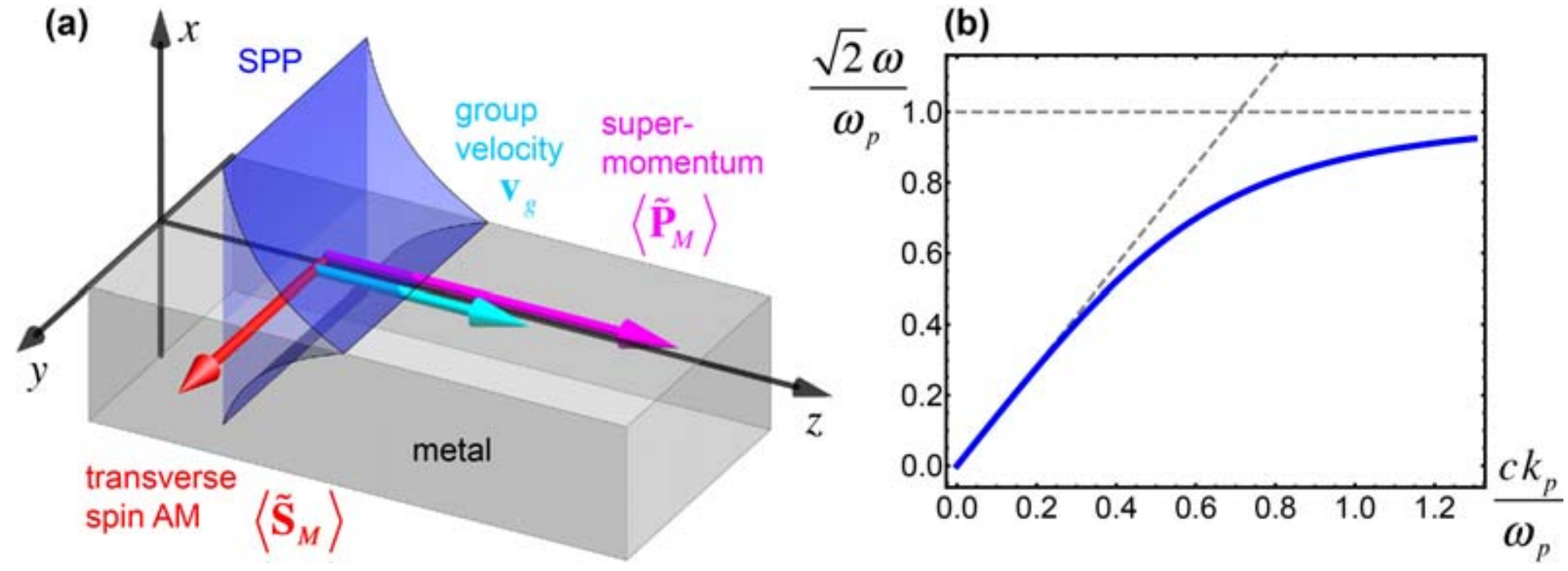

**Fig. 1**. (a) Schematic picture of a surface plasmon-polariton (SPP) wave at the metal-vacuum interface [4,17]. The subluminal group velocity, super-momentum (see Ref. [17]), and the transverse spin (27) are schematically shown. (b) The dispersion of the SPP $\omega(k_p)$ obtained from Eqs. (21) and (25).

The electric and magnetic fields of a SPP wave are described by equations [4,16,17]:

$$\mathbf{E} = A \begin{cases} \left( \overline{\mathbf{x}} - i \dfrac{\kappa_1}{k_p} \overline{\mathbf{z}} \right) \exp\left( i k_p z - \kappa_1 x \right), & x > 0 \\ \dfrac{1}{\varepsilon} \left( \overline{\mathbf{x}} + i \dfrac{\kappa_2}{k_p} \overline{\mathbf{z}} \right) \exp\left( i k_p z + \kappa_2 x \right), & x < 0 \end{cases} \tag{23}$$

$$\mathbf{H} = A \begin{cases} \overline{\mathbf{y}} \dfrac{k_0}{k_p} \exp\left( i k_p z - \kappa_1 x \right), & x > 0 \\ \overline{\mathbf{y}} \dfrac{k_0}{k_p} \exp\left( i k_p z + \kappa_2 x \right), & x < 0 \end{cases} \tag{24}$$

where $A$ is the field amplitude; the propagation constant $k_p$ and spatial decay constants $\kappa_1$, $\kappa_2$ of the SPP field are

$$k_p = \frac{\sqrt{-\varepsilon}}{\sqrt{-1-\varepsilon}} k_0 , \quad \kappa_1 = \frac{1}{\sqrt{-1-\varepsilon}} k_0 , \quad \kappa_2 = \frac{-\varepsilon}{\sqrt{-1-\varepsilon}} k_0 . \tag{25}$$

The dispersion curve of the SPP following from Eqs. (21) and (25) is shown in Fig. 1b.

Now, substituting (23) and (24) into (17) and (19), with using Eqs. (21) and (25) we readily obtain the canonical momentum distribution

$$\tilde{\mathbf{P}}_M = g|A|^2 \frac{k_0^2}{k_p} \bar{\mathbf{z}} \begin{cases} \dfrac{\varepsilon}{1+\varepsilon} \exp(-2\kappa_1 x), & x > 0 \\ \dfrac{1-\varepsilon+\varepsilon^2}{\varepsilon(1+\varepsilon)} \exp(2\kappa_2 x), & x < 0 \end{cases} \tag{26}$$

(this matches Eqs. (3.4) and (3.9) of Ref. [17]) and the spin density of the SPP wave

$$\tilde{\mathbf{S}}_M = g|A|^2 \bar{\mathbf{y}} \begin{cases} \dfrac{\kappa_1}{k_p} \exp(-2\kappa_1 x) , & x > 0 \\ -\dfrac{\kappa_2}{k_p} \dfrac{2-\varepsilon}{\varepsilon^2} \exp(2\kappa_2 x) , & x < 0 \end{cases} \tag{27}$$

(which corresponds to Eq. (3.13) of [17]). This spin AM is directed oppositely in the vacuum and metal: $S_y < 0$ for $x < 0$, which agrees with the opposite directions of the electric field $\mathbf{E}$ rotation in both media, see Eq. (23). Accordingly, the spin density (27) experiences a "jump" at $x = 0$:

$$\Delta\tilde{\mathbf{S}}_M = \tilde{\mathbf{S}}_M(x>0) - \tilde{\mathbf{S}}_M(x<0) = g|A|^2 \frac{\kappa_1}{k_p} \frac{2(\varepsilon-1)}{\varepsilon} \bar{\mathbf{y}} = 2g|A|^2 \frac{1-\varepsilon}{(-\varepsilon)^{3/2}} \bar{\mathbf{y}} . \tag{28}$$

This is in contrast to the "naïve" Minkowski spin (9) that is continuous at the interface:

$$\mathbf{S}_M = g|A|^2 \bar{\mathbf{y}} \begin{cases} \dfrac{\kappa_1}{k_p} \exp(-2\kappa_1 x) , & x > 0 \\ -\dfrac{\kappa_2}{k_p \varepsilon} \exp(2\kappa_2 x) , & x < 0 \end{cases} . \tag{29}$$

Formula (27) provides an adequate description of the SPP transverse spin predicted earlier [24], and correctly characterizes the total spin of the SPP proportional to $\int_{-\infty}^{\infty} \tilde{\mathbf{S}}_M(x)\, dx$ [17]. Now we use Eq. (27) for evaluation of the spin linear momentum. To this end, we note that in the considered SPP geometry (Fig. 1a), all types of spin are *y*-directed and *z*-independent, and Eq. (13) simplifies to

$$\nabla \times \left( \overline{\mathbf{y}} S_y \right) = \overline{\mathbf{z}} \frac{\partial S_y}{\partial x} \tag{30}$$

which in application to Eq. (27) results in

$$\tilde{\mathbf{P}}_M^S = g\left|A\right|^2 \frac{1-\varepsilon}{\left(-\varepsilon\right)^{3/2}} \overline{\mathbf{z}} \delta\left(x\right) + g\left|A\right|^2 \frac{k_0^2}{k_p} \overline{\mathbf{z}} \begin{cases} \dfrac{1}{1+\varepsilon} \exp\left(-2\kappa_1 x\right), & x > 0 \\ \dfrac{2-\varepsilon}{1+\varepsilon} \exp\left(2\kappa_2 x\right), & x < 0 \,. \end{cases} \tag{31}$$

It is helpful to compare this result with the "dispersion-free" spin momentum $\mathbf{P}_M^S$ that follows from (13), (30) and (7):

$$\mathbf{P}_M^S = \frac{1}{2} \nabla \times \mathbf{S}_M = g\left|A\right|^2 \frac{k_0^2}{k_p} \overline{\mathbf{z}} \begin{cases} \dfrac{1}{1+\varepsilon} \exp\left(-2\kappa_1 x\right), & x > 0 \\ \dfrac{\varepsilon}{1+\varepsilon} \exp\left(2\kappa_2 x\right), & x < 0 \end{cases} \tag{32}$$

The difference between (31) and (32) only exists in the metal and at the interface ( $x \le 0$ ) and can be written as

$$\Delta \mathcal{P} = \frac{1}{2} \nabla \times \left( \tilde{\mathbf{S}}_M - \mathbf{S}_M \right) = \tilde{\mathbf{P}}_M^S - \mathbf{P}_M^S = \left( \Delta \mathcal{P} \right)^{\text{surf}} + \left( \Delta \mathcal{P} \right)^{\text{vol}} \tag{33}$$

with explicitly separated surface (singular) and volume contributions:

$$\left( \Delta \mathcal{P} \right)^{\text{surf}} = g\left|A\right|^2 \frac{1-\varepsilon}{\left(-\varepsilon\right)^{3/2}} \overline{\mathbf{z}} \delta\left(x\right), \quad \left( \Delta \mathcal{P} \right)^{\text{vol}} = -g\left|A\right|^2 \frac{k_0^2}{k_p} \overline{\mathbf{z}} \frac{2\left(1-\varepsilon\right)}{-1-\varepsilon} e^{2\kappa_2 x}, \; x < 0 \,. \tag{34}$$

The term with delta-function appears due to the spin AM discontinuity (27), (28) whereas the volume part of (34) describes the additional momentum contribution that was "lost" in the phenomenological SPP analysis in Section 3 of Ref. [17] (but "found" in the microscopic approach presented in subsequent sections of [17], which will be demonstrated in the next Section of this paper).

Now we briefly discuss some aspects of the new results (31) and (33), (34). First to note, with allowance for Eqs. (25), the delta-function term (34) guarantees the zero value for the "total" additional momentum (33) of the SPP cross section

$$\langle \Delta \mathcal{P} \rangle = \int_{-\infty}^{\infty} \Delta \mathcal{P} dx = 0 \tag{35}$$

as well as for the integral spin momentum

$$\langle \tilde{\mathbf{P}}_M^S \rangle = \int_{-\infty}^{\infty} \tilde{\mathbf{P}}_M^S dx = 0$$

– which is required by the general theory [22,23,26] and is associated with the divergence-less nature of the quantities (20), (33). Second, correction (33) of the spin momentum (from (32) to (31)) is equivalent to transition from (14) to (14a), i.e. to the attachment of the second line of Eq. (14a) to the Philbin's kinetic Minkowski momentum (14). Therefore, the "true" kinetic momentum of the SPP is expressed by relation

$$\tilde{\mathcal{P}}_M = \tilde{\mathbf{P}}_M + \tilde{\mathbf{P}}_M^S = \tilde{\mathcal{P}}_M^P + \Delta \mathcal{P} = (\Delta \mathcal{P})^{\text{surf}} + \tilde{\mathcal{P}}_M^{\text{vol}} \tag{36}$$

(see Eqs. (33) and (34)) and appears to be singular, due to first Eq. (34). In the relation (36), $\tilde{\mathcal{P}}_M^P$ corresponds to the Philbin's dispersive-medium momentum (14)

$$\tilde{\mathcal{P}}_M^P = g|A|^2 \frac{k_0^2}{k_p} \bar{\mathbf{z}} \begin{cases} \exp(-2\kappa_1 x), & x > 0 \\ \dfrac{1-\varepsilon+2\varepsilon^2}{\varepsilon(1+\varepsilon)} \exp(2\kappa_2 x), & x < 0 \end{cases}, \tag{37}$$

which expectedly presents the same result as Eq. (3.10) of [17], and

$$\tilde{\mathcal{P}}_M^{\text{vol}} = g|A|^2 \frac{k_0^2}{k_p} \bar{\mathbf{z}} \begin{cases} \exp(-2\kappa_1 x), & x > 0 \\ \dfrac{1}{\varepsilon} \exp(2\kappa_2 x), & x < 0 \end{cases}. \tag{38}$$

Remarkably, expression (38) coincides with the kinetic Abraham momentum $\mathcal{P}_A$ of the SPP obtained in Eq. (3.7) of [17] without any account for dispersion:

$$\mathcal{P}_A = \tilde{\mathcal{P}}_M^{\text{vol}}. \tag{39}$$

This is an interesting conclusion, and it suggests that the dispersion-modified kinetic Minkowsky momentum can be equivalent to the dispersion-free Abraham momentum as an instrument for description of the energy flow and group velocity [16,17]. However, Eq. (39) is associated with the special form of the SPP field, adopted in this paper, in particular, with the simple model of the metal permittivity (21), (22), and can hardly be generalized to other cases. More instructive and demonstrative are the singular terms in (31), (33), (36), and in the next Section we consider their physical nature via the microscopic analysis.

**4. Microscopic approach to the SPP momentum**

Here we briefly consider how the modifications of the SPP momentum description brought about by the new definitions of the field momentum in a dispersive medium (Eqs. (20) and (14a), (31), (36), (38)) are compatible with the microscopic analysis. Following Refs. [16,17], the microscopic approach is based on the separation of the microscopic electromagnetic field ($\mathbf{E}$, $\mathbf{H}$) and charges/currents inside the medium. The metal is described by the Bloch hydrodynamic model for electron plasma, in which the electron density is characterized by the uniform "background" density $n_0$ (see Eq. (22)) modified by small additive time-harmonical perturbation, $\mathrm{Re}\left[\tilde{n}(\mathbf{r})\exp(-i\omega t)\right]$, and the local velocity of electrons is taken in the form $\mathrm{Re}\left[\tilde{\mathbf{v}}(\mathbf{r})\exp(-i\omega t)\right]$. Then the free-space Maxwell equations with $\varepsilon = \mu = 1$ and the densities of charge $e\tilde{n}$ and current $en_0\tilde{\mathbf{v}}$,

$$\nabla\cdot\mathbf{H} = 0\,,\quad \mathbf{H} = -\frac{i}{k_0}\nabla\times\mathbf{E}\,,\quad \nabla\cdot\mathbf{E} = 4\pi e\tilde{n}\,,\quad \mathbf{E} = \frac{i}{k_0}\nabla\times\mathbf{H} - i\frac{4\pi e n_0}{\omega}\tilde{\mathbf{v}}\,, \tag{40}$$

together with the hydrodynamic electron-gas equation [17]

$$-i\omega n_0 m\tilde{\mathbf{v}} = n_0 e\mathbf{E} - m\beta^2\nabla\tilde{n}\,,$$

yield for the medium 2 ($x < 0$):

$$\mathbf{E} = \frac{A}{\varepsilon}\left\{\left[-(1-\varepsilon)e^{\gamma x} + e^{\kappa_2 x}\right]\overline{\mathbf{x}} + i\left[-(1-\varepsilon)\frac{k_p}{\gamma}e^{\gamma x} + \frac{\kappa_2}{k_p}e^{\kappa_2 x}\right]\overline{\mathbf{z}}\right\}\exp\left(ik_p z\right)\,,\quad x<0\,, \tag{41}$$

(the magnetic field is still described by Eq. (24)),

$$\tilde{n} = \frac{A}{4\pi e}\frac{\varepsilon-1}{\varepsilon}\left(\gamma - \frac{k_p^2}{\gamma}\right)e^{\gamma x}\exp\left(ik_p z\right),\quad x<0\,, \tag{42}$$

$$\tilde{\mathbf{v}} = i\frac{A}{\varepsilon}\frac{e}{m\omega}\left[\left(-e^{\gamma x} + e^{\kappa_2 x}\right)\overline{\mathbf{x}} + i\left(-\frac{k_p}{\gamma}e^{\gamma x} + \frac{\kappa_2}{k_p}e^{\kappa_2 x}\right)\overline{\mathbf{z}}\right]\exp\left(ik_p z\right),\quad x<0\,. \tag{43}$$

Here $\gamma^2 = k_p^2 - \varepsilon\,\omega^2/\beta^2$ where the coefficient $\beta^2 = (3/5)v_F^2$, involving the Fermi velocity of electrons $v_F$, is responsible for the additional quantum pressure; $\varepsilon$ is still described by (21) although in (41) – (43), like everywhere in this Section, it is not postulated but derived from the microscopic analysis. According to [16,17], in further consideration we imply the limit $\beta^2 \to 0$ and, correspondingly, $\gamma \to \infty$.

In the limit $\gamma \to \infty$, the $\gamma$-containing terms have non-zero values only in the closest vicinity of the interface in the metal half-space $x < 0$ so we will call them "near-surface terms". Their contributions seem to be negligible but we explicitly hold them because they are crucial for fulfillment of the boundary conditions (continuous electric field and zero normal velocity of electrons (43) at $x = 0$) and additionally characterize the near-surface behavior of the field characteristics. Besides, in some cases such terms provide specific non-vanishing near-surface contributions due to the limiting transition

$$\gamma \exp\left(\gamma x\right) \to \delta\left(x\right) . \tag{44}$$

Now let us consider the momentum calculation in the metal. According to Refs. [8–10], the electromagnetic momentum includes the field and material contributions, and the field contribution is described by the Poynting vector of the free-space field of Eqs. (40)

$$\mathcal{P}_0 = g k_0 \operatorname{Re}\left(\mathbf{E}^* \times \mathbf{H}\right) . \tag{45}$$

In view of Eqs. (40), its spin-orbital decomposition reads (cf. Eqs. (7) – (9))

$$\mathcal{P}_0 = \underbrace{\frac{g}{2} \operatorname{Im}\left[\mathbf{E}^* \cdot \left(\nabla\right) \mathbf{E} + \mathbf{H}^* \cdot \left(\nabla\right) \mathbf{H}\right] - 2\pi g \frac{n_0 e}{c} \operatorname{Im}\left(\mathbf{H}^* \times \tilde{\mathbf{v}}\right)}_{\text{canonical}}$$

$$+ \underbrace{\frac{g}{4} \nabla \times \operatorname{Im}\left[\left(\mathbf{E}^* \times \mathbf{E}\right) + \left(\mathbf{H}^* \times \mathbf{H}\right)\right] - 2\pi g e \operatorname{Im}\left(\mathbf{E}^* \tilde{n}\right)}_{\text{spin}} . \tag{46}$$

The material contribution is calculated considering a long but finite wave packet and the cycle-averaged force density acting on the dipole moment induced in the medium by an external electromagnetic field [10,17]. Afterwards, the length of the wave packet tends to infinity with the result

$$\mathcal{P}_{\text{mat}} = \frac{g\,\omega}{2} \frac{d\varepsilon}{d\omega} \operatorname{Im}\left[\mathbf{E}^* \cdot \left(\nabla\right) \mathbf{E}\right] + \left(\varepsilon - 1\right) g k_0 \operatorname{Re}\left(\mathbf{E}^* \times \mathbf{H}\right) .$$

In combination with (45) this gives the kinetic momentum that corresponds to the Philbin's expression (14)

$$\tilde{\mathcal{P}}_M^P = \varepsilon \underbrace{g k_0 \operatorname{Re}\left(\mathbf{E}^* \times \mathbf{H}\right)}_{\mathcal{P}_0} + \frac{g\omega}{2} \frac{d\varepsilon}{d\omega} \operatorname{Im}\left[\mathbf{E}^* \cdot \left(\nabla\right) \mathbf{E}\right] . \tag{47}$$

In application to the SPP field (41), it reads

$$\tilde{\mathcal{P}}_M^P = g\left|A\right|^2 \frac{k_0^2}{k_p} \overline{\mathbf{z}} \left[ -\frac{1-\varepsilon}{\varepsilon\left(1+\varepsilon\right)} e^{\left(\kappa_2 + \gamma\right)x} + \frac{1-\varepsilon+2\varepsilon^2}{\varepsilon\left(1+\varepsilon\right)} e^{2\kappa_2 x} \right], \quad x < 0 , \tag{48}$$

which, excluding the near-surface term, coincides with the phenomenological expression (37). The canonical and spin parts of this momentum follow from decomposition of the $\mathcal{P}_0$ term in (47) with using Eq. (46). Then, with the help of Eqs. (A1) and (A2) (see Appendix), we obtain

$$\tilde{\mathbf{P}}_M = g\left|A\right|^2 \frac{k_0^2}{k_p}\bar{\mathbf{z}}\left[\frac{\left(1-\varepsilon\right)^2}{2\left(1+\varepsilon\right)}e^{2\gamma x} - \frac{\left(1-\varepsilon\right)\left(2-\varepsilon\right)}{2\varepsilon}e^{\left(\gamma+\kappa_2\right)x} + \frac{1-\varepsilon+\varepsilon^2}{\varepsilon\left(1+\varepsilon\right)}e^{2\kappa_2 x}\right], \quad x<0 \tag{49}$$

and

$$\mathbf{P}_M^S = g\left|A\right|^2 \frac{k_0^2}{k_p}\bar{\mathbf{z}}\left[\frac{\left(1-\varepsilon\right)^2}{2\left(1+\varepsilon\right)}\left(-e^{\gamma x}+e^{\kappa_2 x}\right)e^{\gamma x} + \frac{\varepsilon}{1+\varepsilon}e^{2\kappa_2 x}\right], \quad x<0. \tag{50}$$

Both results well correspond to the macroscopic equations (26) and (32). Note that due to the near-surface terms in (49), the microscopic canonical momentum is continuous at the interface $x = 0$ (cf. Eq. (26)).

To find further momentum constituents, let us address microscopically the spin AM of the SPP field. Keeping in mind that the spin contains only the $y$-component, Eq. (30) gives a direct way to the spin constituent associated with the spin momentum (50)

$$\mathbf{S}_M = g\left|A\right|^2 \frac{k_0^2}{k_s}\bar{\mathbf{y}}\frac{1}{\varepsilon+1}\left(-\frac{\left(1-\varepsilon\right)^2}{2\gamma}e^{2\gamma x} + \frac{\left(1-\varepsilon\right)^2}{\gamma+\kappa_2}e^{\left(\gamma+\kappa_2\right)x} + \frac{\varepsilon}{\kappa_2}e^{2\kappa_2 x}\right), \quad x<0, \tag{51}$$

which upon the condition $\gamma \to \infty$ reduces to (29). This is the "naïve" Minkowski spin that, according to Eq. (4.21) of [17] equals to

$$\mathbf{S}_M = \frac{g}{2}\mathrm{Im}\left(\varepsilon\,\mathbf{E}^* \times \mathbf{E}\right). \tag{52}$$

There exists another spin constituent that is associated with the elliptic motion of free electrons driven by the rotating electric field (23) or (41) (see Fig. 2). It has been considered in Ref. [17] and described by equation

$$\mathbf{S}_{\mathrm{mat}} = \frac{n_0 m}{2\omega}\mathrm{Im}\left(\tilde{\mathbf{v}}^* \times \tilde{\mathbf{v}}\right) = \frac{n_0 e^2}{2m\omega^3}\mathrm{Im}\left(\mathbf{E}^* \times \mathbf{E}\right) = \frac{g\,\omega}{2}\frac{d\varepsilon}{d\omega}\mathrm{Im}\left(\mathbf{E}^* \times \mathbf{E}\right). \tag{53}$$

This "material" spin corresponds to the dispersion terms in (15), and the sum $\tilde{\mathbf{S}}_M = \mathbf{S}_M + \mathbf{S}_{\mathrm{mat}}$ forms the true dispersion-modified spin (19) for the SPP. Therefore, to finalize the SPP momentum calculation, we should determine the quantity whose macroscopic prototype is presented by Eq. (33):

$$\Delta\mathcal{P} = \frac{1}{2}\nabla \times \left(\tilde{\mathbf{S}}_M - \mathbf{S}_M\right) = \frac{1}{2}\nabla \times \mathbf{S}_{\mathrm{mat}}.$$

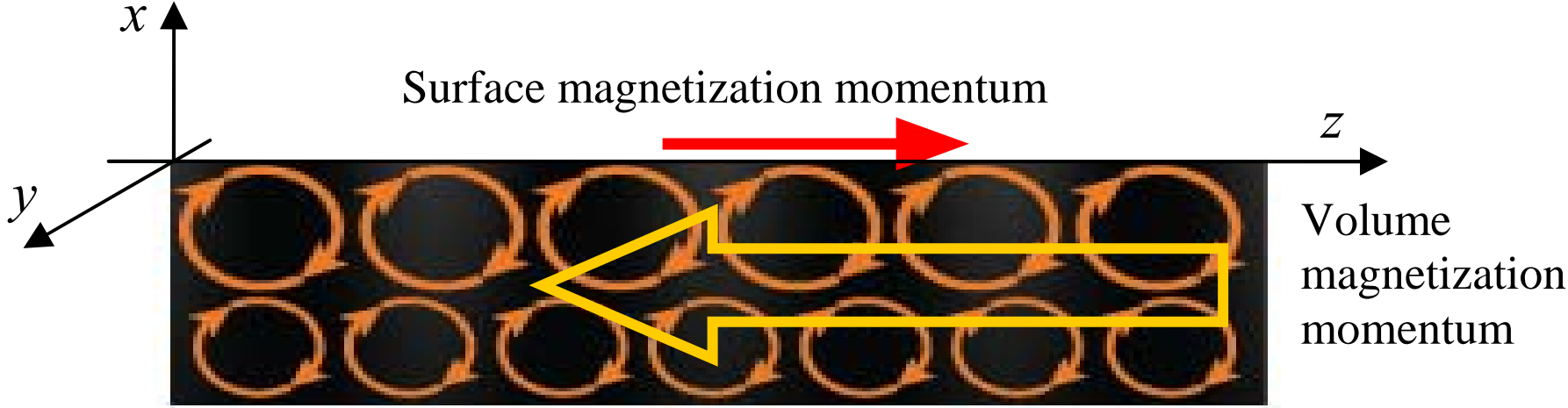

**Fig. 2.** Illustration of the elliptic trajectories of electrons and the magnetization momentum formation. The size of the ellipses and the electron velocities decay with the off-surface distance in the half-space $x < 0$.

The evaluation is straightforward with the known distributions of the SPP field (41) and electron velocity (43). But the different expressions (53) are not fully equivalent; the first equality is more accurate because in further transformations, the simple proportionality between $\tilde{\mathbf{v}}$ and $\mathbf{E}$ was supposed with discarding the near-surface terms [17]. Therefore we use the "original" form

$$\Delta\mathcal{P} = \frac{n_0 m}{4\omega}\operatorname{Im}\left[\nabla\times\left(\tilde{\mathbf{v}}^*\times\tilde{\mathbf{v}}\right)\right] \tag{54}$$

which, based on Eq. (43), can be transformed by the procedure described in the Appendix and eventually yields, for $x \le 0$,

$$\mathbf{S}_{\text{mat}} = -g\left|A\right|^2 \frac{2\kappa_2}{k_p}\overline{\mathbf{y}}\frac{1-\varepsilon}{\varepsilon^2}\left[e^{2\kappa_2 x} - e^{(\kappa_2+\gamma)x}\right], \tag{55}$$

$$\Delta\mathcal{P} = -g\left|A\right|^2 \frac{1-\varepsilon}{\varepsilon^2}\overline{\mathbf{z}}\left[\frac{2\kappa_2^2}{k_p}e^{2\kappa_2 x} - \frac{\kappa_2}{k_p}\delta(x)\right]$$

$$= -g\left|A\right|^2 \frac{k_0^2}{k_p}\overline{\mathbf{z}}\frac{2(1-\varepsilon)}{-1-\varepsilon}e^{2\kappa_2 x} + g\left|A\right|^2\frac{1-\varepsilon}{(-\varepsilon)^{3/2}}\overline{\mathbf{z}}\delta(x),$$

that is, precisely the same result as (33), (34). Therefore, the microscopic model of the SPP perfectly confirms the results obtained with the general phenomenological approach in Section 3.

As a final remark, we emphasize that the momentum $\Delta\mathcal{P}$ (more exactly, its volume part $(\Delta\mathcal{P})^{\text{vol}}$) is actually present in Refs. [16,17], although not deduced from the general scheme of the momentum description in complex media. It appears in the analysis of the newly predicted effect of the metal magnetization due to rotational motion of the free electrons (Section 4.4 of [17]) and is interpreted as the "magnetization momentum" $\mathcal{P}_{\text{magn}}$; one can see that its expression (4.29) of Ref. [17] completely coincides with $(\Delta\mathcal{P})^{\text{vol}}$ (34).

This means that due to the corrected expressions for the field momentum (16a) and (14a), the magnetization momentum finds its place in the unified picture of the field momentum and AM in complex media. Really, obeying the easily verifiable relation $\mathcal{P}_{\text{magn}} = \frac{1}{2}\nabla \times \mathbf{S}_{\text{mat}}$, it appears as a part of the linear spin momentum. Now, its immediate relation with the vortex motion of electrons discloses the general physical mechanism of the spin momentum genesis and the deep analogies with similar phenomena in electromagnetism, fluid mechanics, etc. [29–34] where a linear macroscopic current emerges in the system of inhomogeneously distributed microscopic vorticities. In our case, the volume magnetization appears due to the incomplete compensation of oppositely directed electron velocities in adjacent horizontal layers of the metal (Fig. 2) and is proportional to the "vorticity gradient" ($dS_{\text{mat}} / dx$). Additionally, at the interface, $S_{\text{mat}}$ abruptly changes to zero (this fact is seen from the second term in parentheses of Eq. (55) which, at large $\gamma$, is zero almost everywhere in the volume but rapidly increase to 1 at $x = 0$), which corresponds to a delta-like gradient but of the opposite sign. Therefore, the surface part of expression (34) $\left(\Delta\mathcal{P}\right)^{\text{surf}}$ should also be considered as a part of the magnetization momentum, in addition to the volume part discussed in [16,17] (see Fig. 2). Accordingly, the true form of the magnetization momentum is $\mathcal{P}_{\text{magn}} = \Delta\mathcal{P}$ (34) with the singular part $\mathcal{P}_{\text{magn}}^{\text{surf}} = \left(\Delta\mathcal{P}\right)^{\text{surf}}$.

This surface part of the magnetization momentum is similar to the surface Ampere current in magnets [32,33]. Noteworthy, according to (35), the integral magnetization momentum over the whole SPP cross section is $\left\langle \mathcal{P}_{\text{magn}} \right\rangle = \int_{-\infty}^{\infty} \Delta\mathcal{P}\, dx = 0$ so that the surface (singular) current forms a "closed circuit" with the volume (distributed) part.

**5. Conclusion**

The main result of this work is the unified description of the momentum and angular momentum (AM) in lossless dispersive media supplied by Eqs. (16a) and (17) – (20). Due to enhanced interpretation of the known relation (13) and recognition that every sort of the spin AM is accompanied by the corresponding linear spin momentum, we accomplish the recently proposed scheme [16,17] to its logical end. Now the system of equations for all the constituents: canonical (orbital) and spin linear momenta, orbital and spin AMs, appears in the perfect form, neatly and concisely including the dispersion corrections in the same manner as in the well-known Brillouin's relation (2), (3) for the energy. Importantly, all other conclusions of [16,17] relating the physical interpretation of the kinetic and canonical pictures, meaningful discrepancies and appropriateness of the Abraham-type and Minkowski-type paradigms as well as predicted novel effects in the SPP physics remain unchanged.

The refined general prescriptions are applied to the SPP case which provides an example of the highly structured field in strongly inhomogeneous and essentially dispersive medium, enabling the consistent and meaningful microscopic analysis. This microscopic analysis is performed, and it completely supports expectations based on the phenomenological ground. Additionally, the magnetization momentum which was introduced in [16,17] upon considering the special effect of the SPP-induced magnetization, has been included into the unified general scheme and appears to be its essential part. Moreover, its singular component associated with the surface current is revealed and explained on the footing of far-reaching electromagnetic and hydrodynamic analogies.

The microscopic analysis is based on the free-electron-gas model of the metal but partly includes the quantum pressure influence immanent in the hydrodynamic model of the electron plasma. This influence is supposed to be negligible but some residual effects are taken into account in the form of corrections to the distributions of the electric field, electron density and velocity, sharply localized near the metal-vacuum interface. These "near-surface" terms do not affect the main volume properties of the SPP wave but provide conceptually meaningful contributions necessary for fulfillment of the boundary conditions. Besides, they describe some principal details of the near-interface behavior of various momentum components. In particular, it is these terms that stipulate the singular surface part of the magnetization momentum (34); another remarkable observation is that the canonical momentum of the SPP field appears to be continuous at the interface (cf. Eqs. (49) and (26)). At the same time, the meaning and consequences of the near-surface terms needs additional elucidation and, probably, will require a more accurate model of the electron properties of metal, which is a prospective direction for further development.

We hope that the present work provides a suitable and efficient toolkit for analysis and description of the momentum and AM of light in dispersive and inhomogeneos (but isotropic and lossless) media. It can be used in a variety of modern problems, involving photonic crystals, metamaterials, and optomechanical systems.

**Appendix.**

We consider transformations of the second line of Eq. (46). In the first term, $\mathbf{H}^* \times \mathbf{H}$ for the SPP field (24) vanishes and we must calculate $\frac{g}{4}\nabla \times \operatorname{Im}\left(\mathbf{E}^* \times \mathbf{E}\right)$. By substitution of (41) we find

$$\frac{g}{4}\operatorname{Im}\left(\mathbf{E}^* \times \mathbf{E}\right) = -\frac{g}{2}\overline{\mathbf{y}}\operatorname{Im}\left(E_x^* E_z\right)$$

$$= -\frac{g|A|^2}{2\varepsilon^2}\overline{\mathbf{y}}\left[\left(1-\varepsilon\right)^2 \frac{k_p}{\gamma} e^{2\gamma x} + \frac{\kappa_2}{k_p} e^{2\kappa_2 x} - \left(1-\varepsilon\right)\left(\frac{\kappa_2}{k_p} + \frac{k_p}{\gamma}\right) e^{(\kappa_2+\gamma)x}\right].$$

Further, direct application of the rule (30) leads to representation

$$
\begin{aligned}
&\frac{g}{4}\operatorname{Im}\nabla\times\left(\mathbf{E}^{*}\times\mathbf{E}\right)\\
&=-\frac{g\left|A\right|^{2}}{2\varepsilon^{2}}\bar{\mathbf{z}}\left[2k_{p}\left(1-\varepsilon\right)^{2}e^{2\gamma x}+\frac{2\kappa_{2}^{2}}{k_{p}}e^{2\kappa_{2}x}-\left(1-\varepsilon\right)\left(\frac{\kappa_{2}}{k_{p}}+\frac{k_{p}}{\gamma}\right)\left(\kappa_{2}+\gamma\right)e^{\left(\kappa_{2}+\gamma\right)x}\right]\\
&\xrightarrow[\gamma\to\infty]{}-\frac{g\left|A\right|^{2}}{2\varepsilon^{2}}\bar{\mathbf{z}}\left[2k_{p}\left(1-\varepsilon\right)^{2}e^{2\gamma x}-\left(1-\varepsilon\right)\left(\frac{\kappa_{2}^{2}}{k_{p}}+k_{p}\right)e^{\left(\kappa_{2}+\gamma\right)x}\right.\\
&\qquad\left.+\frac{2\kappa_{2}^{2}}{k_{p}}e^{2\kappa_{2}x}-\left(1-\varepsilon\right)\frac{\kappa_{2}}{k_{p}}\delta\left(x\right)\right]
\end{aligned}
\tag{A1}
$$

where the limit transition $\gamma\to\infty$ is performed and Eqs. (44) and (25) are employed.

In the second term of the second line of Eq. (46), substitution of (41) and (42) with the same limit transition and application of Eq. (44) gives

$$
\begin{aligned}
-2\pi ge\operatorname{Im}\left(\mathbf{E}^{*}\tilde{n}\right)&=\frac{g}{2}\left|A\right|^{2}\frac{1-\varepsilon}{\varepsilon^{2}}\left[\left(1-\varepsilon\right)k_{p}\left(1-\frac{k_{p}^{2}}{\gamma^{2}}\right)e^{2\gamma x}-\frac{\kappa_{2}}{k_{p}}\gamma e^{\kappa_{2}x+\gamma x}+\frac{\kappa_{2}k_{p}}{\gamma}e^{\kappa_{2}x+\gamma x}\right]\\
&\xrightarrow[\gamma\to\infty]{}\frac{g}{2}\left|A\right|^{2}\frac{1-\varepsilon}{\varepsilon^{2}}\left[\left(1-\varepsilon\right)k_{p}e^{2\gamma x}-\frac{\kappa_{2}}{k_{p}}\delta\left(x\right)\right].
\end{aligned}
\tag{A2}
$$

Note that the terms with delta-functions in (A1) and (A2) mutually cancel. Transformation of Eq. (54) is performed similarly to (A1) with taking Eqs. (43) instead of (41).

## Acknowledgements

This work was supported by the Ministry of Education and Science of Ukraine, Project No. 582/18, RIKEN iTHES Project, MURI Center for Dynamic Magneto-Optics via the AFOSR Award No. FA9550-14-1-0040, the Japan Society for the Promotion of Science (KAKENHI), the IMPACT program of JST, CREST grant No. JPMJCR1676, the John Templeton Foundation, the RIKEN-AIST “Challenge Research” program, and the Australian Research Council.